\documentclass[sn-mathphys,Numbered]{sn-jnl}

\usepackage{graphicx}%
\usepackage{multirow}%
\usepackage{amsmath,amssymb,amsfonts}%
\usepackage{amsthm}%
\usepackage{mathrsfs}%
\usepackage[title]{appendix}%
\usepackage{xcolor}%
\usepackage{textcomp}%
\usepackage{manyfoot}%
\usepackage{booktabs}%
\usepackage{algorithm}%
\usepackage{algorithmicx}%
\usepackage{algpseudocode}%
\usepackage{listings}%
\usepackage[caption=false,font=normalsize,labelfont=sf,textfont=sf]{subfig}

\theoremstyle{thmstyleone}%
\theoremstyle{thmstyletwo}%

\theoremstyle{thmstylethree}%

\begin{document}

\title[Hybrid CNN Bi-LSTM neural network for Hyperspectral image classification]{Hybrid CNN Bi-LSTM neural network for Hyperspectral image classification}


\author*[1]{\fnm{Alok Ranjan } \sur{Sahoo}}\email{aloksh90@gmail.com}

\author[2]{\fnm{Pavan } \sur{Chakraborty}}\email{pavan@iiita.ac.in}

\affil*[1]{\orgdiv{Department CSIT}, \orgname{SOA University}, \orgaddress{ \city{Bhubaneswar}, \postcode{751030}, \state{Odisha}, \country{India}}}

\affil[2]{\orgdiv{Department of IT}, \orgname{IIIT Allahabad}, \orgaddress{\street{Jhalwa}, \city{Prayagraj}, \postcode{211015}, \state{UP}, \country{India}}}


\abstract{Hyper spectral images have drawn the attention of the researchers for its complexity to classify. It has nonlinear relation between the materials and the spectral information provided by the HSI image. Deep learning methods have shown superiority in learning this nonlinearity in comparison to traditional machine learning methods. Use of 3-D CNN along with 2-D CNN have shown great success for learning spatial and spectral features. However, it uses comparatively large number of parameters. Moreover, it is not effective to learn inter layer information. Bi-LSTMs are known to learn the correlation among the features in sequential data effectively. Hence, this paper proposes a neural network combining 3-D CNN, 2-D CNN and Bi-LSTM. Joint spatial–spectral features are learnt by the 3-D CNN. The 2-D CNN further learns abstract spatial features. Then, the Bi-LSTM tries to find the class based on the 2-D CNN's output. This combination of 3-D CNN, 2-D CNN and Bi-LSTM uses less the number of trainable parameters. Hence, it reduces the complexity. The performance of this model has been tested on Indian Pines(IP) University of Pavia(PU) and  Salinas Scene(SA) data sets. The results are compared with the state of-the-art deep learning-based models. This model performed better in all three datasets. It could achieve 99.83, 99.98 and 100 percent accuracy using only 30 percent trainable parameters of the state-of-art model in IP, PU and SA datasets respectively. 
}

\keywords{ Hyperspectral image (HSI) classification, Remote sensing, Bi-LSTM, 3-D CNN, 2-D CNN, Spectral–spatial, Deep learning }



\maketitle

\section{Introduction}\label{sec1}

Hyperspectral images(HSI) resemble 3D cubes having spatial and spectral information. Numerous spectral bands covering the same spatial area are frequently present in HSI data, which is useful for identifying different materials. Its pixels are high-dimensional vectors, having  spectral reflectance values from visible to infrared.

Hyperspectral image analysis finds many applications, including precise agriculture\cite{gevaert2015generation}, environmental analysis, military surveillance \cite{shimoni2019hypersectral}, mineral exploration\cite{sudharsan2019survey}, land use and land cover classification \cite{ortac2021comparative}, medical imaging\cite{ul2021review,lu2014medical} etc. The presence of various spectral bands has made its analysis a complex task. 

Initially, the techniques used for classifying hyperspectral images were based on well-known pattern recognition techniques such as support vector machines (SVM) \cite{scholkopf2018learning,bo2015hyperspectral}, multinomial logistic regression \cite{li2010semisupervised,li2011spectral} , K-nearest neighbor classifiers and dynamic or random subspace \cite{du2010random,du2014target}. Linear discriminant analysis (LDA) \cite{bandos2009classification}, Principal component analysis (PCA) \cite{prasad2008limitations,licciardi2011linear} and independent component analysis (ICA) \cite{villa2011hyperspectral} are also used for dimensional reduction and feature extraction. 

Dimensionality reduction was first done and then multinomial logistic regression applied to the HSI dataset by Krishnapuram et al. \cite{krishnapuram2005sparse}. Further, locality adaptive discriminant analysis (LADA) by Wang et al. \cite{wang2017locality}, Multiple Feature Based
Adaptive Sparse Representation (MFASR) \cite{fang2017hyperspectral} and Co-SVM \cite{guo2019hyperspectral} techniques were also developed. However, these models are computationally expensive and suffer from information loss problem \cite{banerjee2023pooled}.   

As CNNs started giving better results a lot of deep learning models \cite{xu2019d,kang2018dual,yu2017unsupervised,song2018hyperspectral,cheng2018exploring} were proposed for this classification problem. Initially, most of them used 2-D CNN for the model. Then, 3D CNN based models \cite{chen2016deep,zhong2017spectral,mou2017unsupervised,paoletti2018deep} were used for it. The main motivation behind using 3-D CNNs was to capture both
spatial and spectral information effectively. However, 3-D-CNN is computationally expensive. If the classes are having having similar textures, then it faces difficulty in performing over many spectral bands. Hence, Roy
et al. proposed a hybrid model \cite{roy2019hybridsn}. They reduced the computational cost by sandwiching  3-D CNN and 2-D CNN. However, the trainable parameters used for it was still high. Moreover, 3D CNN and 2D CNNs are shown to be less effective in learning inter layer information \cite{convlstm}. Further, attempts were made to use LSTM \cite{ZHOU201939}, Bi-LSTM \cite{liu2017bidirectional} for it. LSTM architectures with Transformer along with CNNs were also used by Xu et al. \cite{9900270} and Zhang et al. \cite{zhang2022caevt}.   

It was seen that 3-D CNN can learn joint spatial and spectral features effectively. A combination of 3D and 2D CNN layers (HybridSN \cite{roy2019hybridsn}) could grab both the spectral and spatial features more effectively with less number of parameters. However, it suffers from information loss problem. Inter-layer information could not be grabbed efficiently. From the language models \cite{vyawahare2022pict}, it has been seen that LSTMs could learn correlation between sequence data. However, it fails to grab long term correlation. Hence, Bi-LSTMs are used to solve this problem. Considering the above facts, we propose a simpler model to combine 3-D and 2-D CNN along with Bi-LSTM layer for the classification of HSI data.  We will use the information grabbed by the combined block of 3-D and 2-D CNN to train Bi-LSTM further to give a better accuracy. 

Here, our target is two fold. First, We want to reduce the complexity by using less number of training parameters. Secondly, the model should give better  accuracy than the state-of-art models.

The main contributions of this paper are as follows:

\begin{itemize}
  \item This paper proposes a novel hybrid neural network based on 3-D CNN, 2-D CNN and Bi-LSTM to learn the classes. The Bi-LSTM will use the  spatial and spectral features  learnt by 3-D CNN and 2-D CNN. It will learn the inter layer information.   
  \item Bi-LSTM gets trained by the information from both forward and backward directions. Hence, the inclusion of Bi-LSTM solves the information loss problem and also produces better results. 
  \item The proposed grouping of Bi-LSTM along with with 3D and 2D CNN further decreases the required hyper parameters to be trained. 
\end{itemize}

The paper has been organized as follows:
Section II discusses the basics of Bi-LSTM. Section III proposes hybrid spectral spatial CNN Bi-LSTM (HSSNB). The experimental analysis has been given in Section IV. Further, the conclusion has have been given in Section V.

\section{Bi-LSTM}

RNNs are being used efficiently for many years for various time series data. However, they suffer from vanishing/exploding gradient problems \cite{vanishinggradient}. Given a sequence of data, normal RNNs generally tend to forget past events. They have a bias towards the recent data points. LSTM was introduced to counter these problems. It uses constant error carousel (CEC) to learn long term relationships. Each unit cell maintains the error signal with in itself.  

A vanila LSTM consists of three types of gates:
(1) input gate (2) forget gate (3) output gate. It also include a single cell, block input, peephole connections and an output activation function \cite{vanilalstm,LSTMreview}. The output of this block becomes the input in the next time step. Let us consider $i_{t}$  be the input and and $o_{t}$ be the output at the time step t respectively.\\ 

\textbf{Input block} :
This block basically combines the output of the LSTM unit from the previous time step t-1 and the input of the current time step t.

\begin{equation}
   i^{ib}_{t}= g(W_{ib}i_{t}+R_{ib}o_{t-1}+b_{ib})
\end{equation}

Here, $W_{ib}$ and $R_{ib}$ are the weights corresponding to $i_{t}$ and $o_{t-1}$ respectively. Here, $b_{ib}$ is the bias weight vector. 

\textbf{Input gate} :
Here, the input gate is updated combining the current input at time step t, $i_{t}$, output of the LSTM unit from last time step t-1, $o_{t-1}$ and last step cell value $c_{t-1}$. This step determines the candidate values, $C_{t}$ to be added to the cell state $c_{t}$ and activation values $i^{ac}_{t}$.
\begin{equation}
   i^{ig}_{t}= \sigma(W_{ig}i_{t}+R_{ig}o_{t-1}+p_{ig}\odot c_{t-1}+ b_{ig}) \label{inputgate}
\end{equation}

 $\odot$ means point wise multiplication. Here, $W_{ig}$, $R_{ig}$ and $p_{ig}$ are the weights corresponding to $i_{t}$, $o_{t-1}$ and $c_{t-1}$ respectively. Here, $b_{ig}$ is the bias weight vector. 

\textbf{Forget gate} : Here, the LSTM determines about the elements to be removed from the cell state $c_{t-1}$. This requires current input at the time step t $i_{t}$, output of the LSTM unit from last time step t-1 $o_{t-1}$ and last cell value $c_{t-1}$ as inputs. 

\begin{equation}
   fg_{t}= \sigma(W_{fg}i_{t}+R_{fg}y_{t-1}+p_{fg}\odot c_{t-1}+ b_{fg})
\end{equation}

Here, $W_{fg}$, $R_{fg}$ and $p_{fg}$ are the weights corresponding to $i_{t}$, $o_{t-1}$ and $c_{t-1}$ respectively. Here, $b_{fg}$ is the bias weight vector. 

\textbf{Cell Value}:
To calculate cell value, block input $z_{t}$, input gate value $i_{t}$ and forget gate value $f_{t}$ are combined with last cell value $c_{t-1}$. 

\begin{equation}
   c_{t}= i^{ib}_{t}\odot  i^{ig}_{t} + c_{t-1}\odot fg_{t} \label{Cellvalue}
\end{equation}

\textbf{Output gate}:
Here, the output gate value is calculated using input at the time step t $i_{t}$, output of the LSTM unit from last time step t-1 $o_{t-1}$ and last cell value $c_{t-1}$ as inputs to calculate the output gate value.
\begin{equation}
   o^{og}_{t}= \sigma(W_{og}i_{t}+R_{og}y_{t-1}+p_{og}\odot c_{t-1}+ b_{og})
\end{equation}  

Here, $W_{og}$, $R_{og}$ and $p_{og}$ are the weights corresponding to $i_{t}$, $o_{t-1}$ and $c_{t-1}$ respectively. Here, $b_{og}$ is the bias weight vector.

\textbf{Block output}:
 Here, current cell value $c_{t}$ and output gate value $o_{t}$ are used to find the output.
\begin{equation}
   o_{t}= g(c_{t})\odot o^{og}_{t}
\end{equation}

Here, $\sigma(x)$ =sigmoid function and g(x) = h(x) = hyperbolic tangent function  

\subsection{Back-propagation through time }

To train the model, forward pass as described above is calculated. Then, back propagation through time is calculated to update the weights. Here, the cell value $c_{t}$ receives gradient from the next cell value $c_{t+1}$ and output of the current state $o_{t}$. The deltas are calculated as follows

\begin{equation}
   \delta y_t=\Delta_t+R_{ib}\delta ib_{t+1}+R_{ig}\delta ig_{t+1}+ R_{fg}\delta fg_{t+1} + R_{og}\delta og_{t+1}
\end{equation}

\begin{equation}
   \delta og_t= \delta o_t \odot h(c_t) \odot \sigma^{'}(\overline{og}_t)
\end{equation}

\begin{equation}
   \delta c_t= \delta o_t \odot og_t \odot h^{'}(c_t)+p_{og} \odot \delta og_t + p_{ig} \odot \delta ig_{t+1} + p_{fg} \odot \delta fg_{t+1} + \delta c _{t+1} \odot fg_{t+1}
\end{equation}

\begin{equation}
   \delta fg_t= \delta c_t \odot c_{t-1} \odot \sigma^{'}(\overline{fg}_t)
\end{equation}

\begin{equation}
   \delta i^{ig}_t= \delta c_t \odot i^{ib}_t \odot \sigma^{'}(\overline{i}^{ig}_t)
\end{equation}

\begin{equation}
   \delta i^{ib}_t= \delta c_t \odot i^{ig}_t \odot g^{'}(\overline{i}^{ib}_t)
\end{equation}

Here, $\Delta_t$ is the vector of deltas pass down from the t+1 layer. If we consider $L$ to be the loss function then it would correspond to $\delta E/ \delta og_t$. Here,  $\overline{i}^{ig}_t$, $\overline{og}_t$, $\overline{fg}_t$ and $\overline{i}^{ib}_t$  represent raw values before transformation by the respective transfer functions attached with the input gate, output gate, forget gate and block input respectively.

To train any existing layer below, we need to have delta values for the inputs.   
\begin{equation}
   \delta i_t= \delta W^T_{i^{ib}} \delta i^{ib}_t + \delta W^T_{i^{ig}} \delta i^{ig}_t + \delta W^T_{fg} \delta fg_t + \delta W^T_{og} \delta og_t
\end{equation}

So, the gradients for the weights are calculated as follows. 

\begin{equation}
   \delta W_*=  \sum_{t=0}^{T} <\delta*_t,i_t> 
\end{equation}

\begin{equation}
   \delta p_i=  \sum_{t=0}^{T-1} c_t \odot \delta i^{ib}_{t+1} 
\end{equation}

\begin{equation}
   \delta R_*=  \sum_{t=0}^{T-1} <\delta*_{t+1},o_t> 
\end{equation}

\begin{equation}
   \delta p_{fg}=  \sum_{t=0}^{T-1} c_t \odot \delta fg_{t+1} 
\end{equation}

\begin{equation}
   \delta b_*=  \sum_{t=0}^{T} \delta*_t 
\end{equation}

\begin{equation}
   \delta p_o=  \sum_{t=0}^{T} c_t \odot \delta og_{t} 
\end{equation}

Here, * can be any one of $\overline{i}^{ig}_t$, $\overline{og}_t$, $\overline{fg}_t$, $\overline{i}^{ib}_t$. Here $<*_1,*_2>$ represents  
outer product.
In Bi-LSTM, two LSTMs are applied to the input layer. In forward layer, LSTM gets sequence information in past to future direction. In backward layer, it gets sequence information in future to past direction. It can learn long term dependencies easily using this.

\section{Proposed Network for HSI Classification}

\begin{figure*}
    \centering
    \includegraphics[width=1\textwidth]{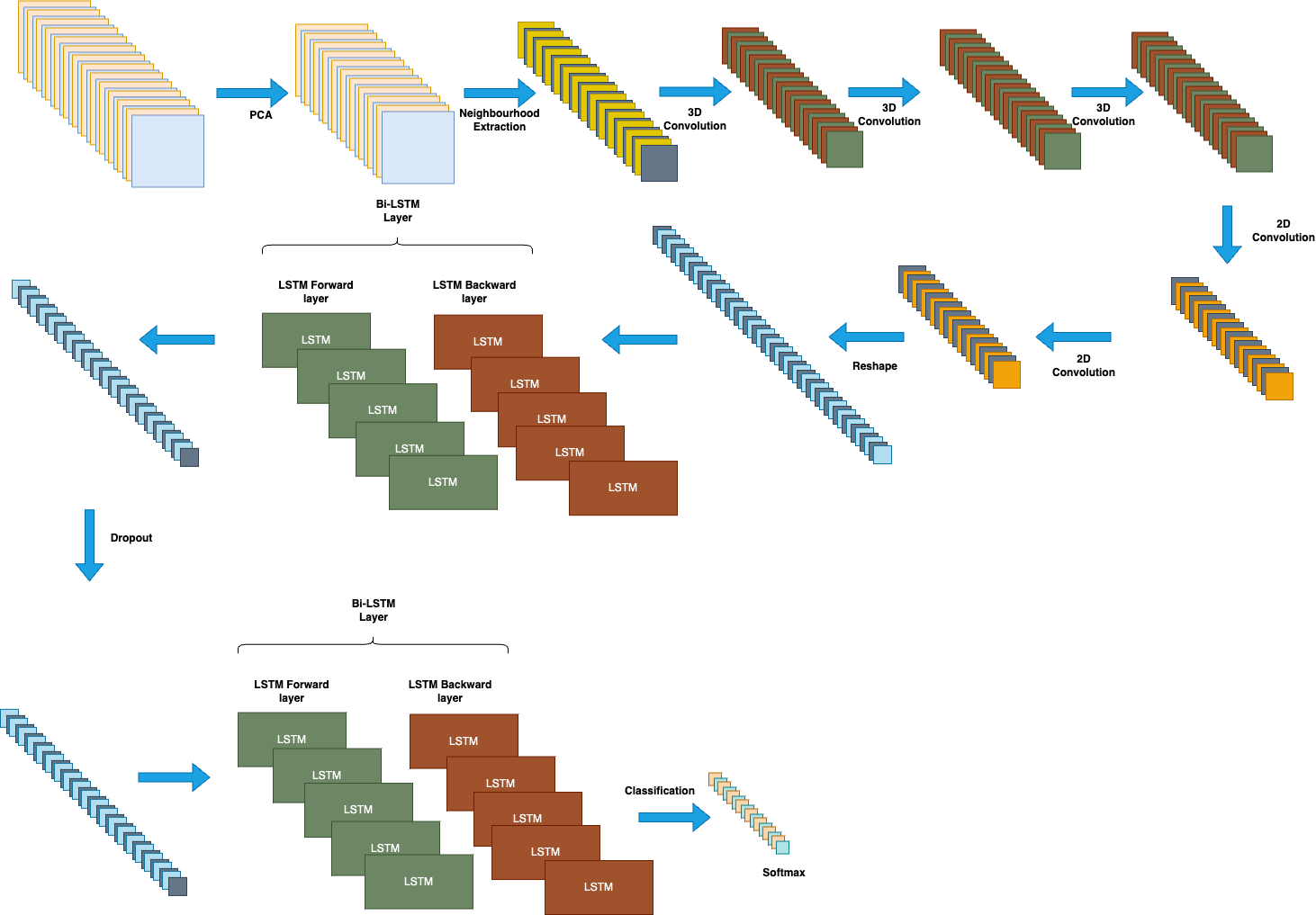}
    \caption{Proposed Network showing the processing of the HSI data}
    \label{proposednetwork}
\end{figure*}
\

This model extends the work of Roy et al.\cite{roy2019hybridsn}. Hybrid-Spectral-Net  outperformed state of art models. However, it used 51,22,176 training parameters, which is significantly higher. Hence, we tried to reduce it and improve the accuracy by incorporating Bi-LSTM layers. LSTMs are preferred over RNNs  due to its effectiveness in reducing the information loss. However, Bi-LSTMs are trained from both forward and backward directions.  Hence, they are more effective to understand the correlations better. Inclusion of Bi-LSTM layer also reduces the required number of training parameters. An additional 2D layer was added for the better understanding of the  abstract level spatial representation.

Let us say that the HSI data cube is represented by   $H \in R^{A \times B \times C}$. Here, A = Width, B= Height and C = No. of spectral bands. 
The pixels of HSI data cube contains C spectral measures. They form one hot label vector which is represented by  $L= (L_1, L_2, L_3,...L_N)$. Here, N = land-cover categories. The exhibition of mixed land-cover classes introduces high inter class variablity and similarity. 
To tackle this problem, the spectral dimension was reduced from C to S using PCA. However, the spatial dimensions were kept unchanged due to its importance in  recognition of the object category. After the dimensionality reduction, the input can be termed as, $I \in R^{A \times B \times S}$, where A = width, B = height and S = reduced number of spectral bands. 3D patches, $P \in R^{D \times D \times S}$, were formed from the data cube. Here $D \times D$ is the spatial window size. The label of the data patch is decided based on the label of the central pixel.

This 3-D data was now convolved over several contiguous bands in the input layer using the 3-D kernel \cite{3dconv}.  It extracts features from the data's spectral and spatial dimensions. The value in the ith feature map of the kth layer at a position (a, b, c) is provided by  

\begin{equation}
f_{k,i}^{a,b,c}=\sigma (b_{k,i}+ \sum_{n}  \sum_{u=0}^{{U_k}-1}\sum_{v=0}^{{V_k}-1} \sum_{g=0}^{{G_k}-1} w_{k,i,m}^{p,q,r} f_{(k-1),m}^{(a+p),(b+q),(c+r)})
\end{equation}
Where, $\sigma$ = activation function, $f_{k,i}^{a,b,c}$ = activation value, $b_{k,i}$ = bias parameter for  ith feature map in the kth layer, $w_{k,i,m}^{p,q,r}$ is the corresponding weight.

Further, the output was reshaped to convolve with the 2D kernel. To cover the complete spatial dimension, striding was used. In the kth layer, the value of (a,b) in the ith feature map is provided by

\begin{equation}
f_{k,i}^{a,b}=\sigma (b_{k,i}+ \sum_{n}  \sum_{u=0}^{{U_k}-1}\sum_{v=0}^{{V_k}-1} w_{k,i,m}^{p,q} f_{(k-1),m}^{(a+p),(b+q)})
\end{equation}
Where, $\sigma$ = activation function, $f_{k,i}^{a,b}$ = activation value, $b_{k,i}$ = bias parameter for  ith feature map in the kth layer, $w_{k,i,m}^{p,q}$ is the corresponding weight. Then, the output from the 2-D CNN was further reshaped to give it as a sequence data to the Bi-LSTM. Fig \ref{proposednetwork} shows the flow of HSI data through the model.

\begin{table}[!t]
\caption{Layer wise summery for Hybrid Bi-LSTM model\label{tab:table1}}
\centering
\begin{tabular}{|c||c||c|}
\hline
Layer(type) & Output Shape & Parameter\\
\hline

$input \textunderscore 1$ (InputLayer) &       ( 25, 25, 30, 1) &  0\\         
                                                                 
 $conv3d \textunderscore 1$ (Conv3D) &          ( 23, 23, 24, 8) &     512 \\       
                                                                 
$ conv3d \textunderscore 2 $(Conv3D)  &      ( 21, 21, 20, 16)  &   5776 \\      
                                                                 
$ conv3d \textunderscore3$ (Conv3D)  &      (19, 19, 18, 32)  &   13856 \\     
                                                                 
 $reshape_1$ (Reshape)  &        (19, 19, 576)   &     0  \\        
                                                                 
 $conv2d_1 (Conv2D)$      &      ( 17, 17, 64)   & 331840    \\ 
                                                                 
 $conv2d_2 (Conv2D)  $      &    (15, 15, 128) &    73856     \\ 
                                                                 
 $reshape_2 (Reshape)$        &  ( 15, 1920)       &    0         \\ 
                                                                 
 $bidirectional_1 (Bidirectional)$ &  (15, 128) & 1016320           \\                                                   
                                                                 
 $dropout_1 (Dropout)$     &     (15, 128)        &    0         \\ 
                                                                 
 $bidirectional_2 (Bidirectional)$ &  (128)      & 98816            \\                                                                                                                 
 $dense_1 (Dense)$        &      (16)              & 2064  \\ 
\hline
Total trainable parameters: 15,43,040 & &\\
\hline
\end{tabular}
\end{table}

\subsection{Dimensions of the proposed model }

The 3d patches received after applying PCA, as explained before, was fed as the input. The first layer was a  3-D CNN layer
having kernel sizes  $3\times3\times7$. 8 filters were used in this layer.  The output from it was further fed to another  3-D CNN layer having kernel size  $3\times3\times5$ and 16 filters. Then, the output was passed through another 3-D CNN layer having kernel size  $3\times3\times3$ and 32 filters. 'Relu' was used as the activation function for them. 

After being reshaped, the output from the preceding 3-D CNN layer was fed through two 2-D CNN layers having kernel size $3\times3$.
 64 and 128 filters were used in these two layers respectively. 'Relu' activation function was used for these two layers as well. Then, it was reshaped to $15\times1920$ and fed to a Bi-LSTM layer (64 output units). After that, the output was passed through a dropout layer (dropout rate = 0.25). The output was again fed to a Bi-LSTM layer (64 output units). The output  of it was finally fed to a softmax layer. The number of classes in the dataset determined the softmax layer's dimension.  Detailed dimensions of the proposed model have been described the table \ref{tab:table1}. Adam optimizer and categorical cross-entropy loss had been applied. The training was done for 100 epochs. 

\section{EXPERIMENTS AND DISCUSSION}

\subsection{Data set description}

To evalute the effectiveness of our model, we used three data sets: Indian Pines (IP), Pavia university (PU) and Salinas (SA). \\

Indian Pines (IP): This data set was collected by AVIRIS sensor. It consists of image of dimension $145\times145 \times 224$. Only 200 out of these 224 spectral bands are considered due to elimination of 24 bands covering water absorption region. The spatial resolution of it is 20m. It has been classified into 16 classes. \\

\begin{table}[!t]
\caption{Indian-pines (IP) dataset \label{IP}}
\centering
\begin{tabular}{|c||c||c|}

\hline
S No. & Class   &  Number of Samples\\      
\hline
1 & Corn & 237 \\
2  &Oats & 20\\
3 & Stone-Steel-Towers &93\\
4 &Alfalfa &46\\

5 & Soybean-clean & 593\\ 
6 & Woods& 1265\\
7 & Buildings-Grass-Trees-Drives & 386\\

8 & Hay-windrowed & 478\\

9 & Wheat & 205\\
10 & Corn-notill &1438\\
11 &Grass-pasture &28\\

12 & Corn-mintill &830\\
13 & Grass-trees & 730\\

14 & Grass-pasture-mowed &28\\     

15 & Soyabean-mintill &2455\\   
16 & Soyabean-notill & 972\\                               
\hline
\end{tabular}
\end{table}

Pavia university (PU): During a fly above Pavia, Italy, the ROSIS sensor collected this data. The image has $610\times610$ pixel dimensions. It has a 1.3 m spatial resolution. It has 103 different spectral bands. It is having 9 different classes.\\


\begin{table}[!t]
\caption{Pavia University (PU) dataset \label{PU}}
\centering
\begin{tabular}{|c||c||c|}

\hline
S No. & Class   &  No. of Samples\\      
\hline

1 & Asphalt & 6631
\\
2 & Bare soil &5029

\\

3 & Self-blocking Bricks & 3682

\\
4 &
Bitumen &1330

\\
5 &Shadows& 947
\\
6 & Meadows & 18649
\\

7 & Trees & 3064

\\
8 & Painted metal sheets & 1345
\\

9 & Gravel & 2099
\\

\hline
\end{tabular}
\end{table}

Salinas scene (SA): It was gathered over Salinas Velly using an AVIRIS sensor with a 224 band. Twenty water-absorption bands were discarded. Pictures having dimensions of $512\times217$ pixels are there. The spatial resolution of it was 3.7 meter. 16 classes are there in its ground truth.

\begin{table}[!t]
\caption{Salinas scene (SA) dataset \label{SA}}
\centering
\begin{tabular}{|c||c||c|}

\hline
S No. & Class   &  No. of Samples\\      
\hline

1 & Fallow smooth & 2678\\
2 & Celery & 3579\\
3 & Fallow plow & 1394\\
4 & Stubble & 3959\\
5 & Fallow & 1946\\
6 & Corn &  3278\\

7 & Soil & 6167\\
8 & Grapes untrained & 11,271\\

9 & Lettuce-4-wk & 1068\\
10 & Vineyard-trellis & 1807\\
11 & Lettuce-5-wk &  1927\\
12 & Vineyard-untrained & 7268\\
13 & Lettuce-6-wk &  916\\
14 & Brocoli green Weed 1 & 2009\\
15 & Lettuce-7-wk & 1070\\
16 & Brocoli green Weed 2&  3726\\

\hline
\end{tabular}
\end{table}








                                                                 

Table \ref{IP},\ref{PU}, \ref{SA} gives the complete details of the above discussed datasets.

\subsection{Experimentation details}

\begin{table}[!t]
\caption{Comparative study \label{comparativestudy}}
\centering
\begin{tabular}{|c||c||c||c||c|}

\hline
Dataset &Models &    Kappa & AA & OA\\      
\hline
IP 
 &2D-CNN  & 87.10 $\pm$ 0.4 & 86.01 $\pm$ 0.9  & 89.19 $\pm$ 0.3\\ 
&3D-CNN & 90.51 $\pm$ 0.4 & 92.07 $\pm$ 0.2 & 91.38 $\pm$ 0.4 \\ 
& SSRN & 98.97 $\pm$ 0.2 & 98.91 $\pm$ 0.5 & 99.15 $\pm$ 0.2\\
& HybridSN & 99.71 $\pm$ 0.1 & 99.63 $\pm$ 0.2 & 99.75 $\pm$ 0.1\\
& HSSNB & 99.80 $\pm$ 0.1 & 99.89 $\pm$ 0.1 & 99.83 $\pm$ 0.1\\
\hline

\hline
 PU 
 &2D-CNN  & 97.10 $\pm$ 0.5 & 96.51 $\pm$ 0.2  & 97.70 $\pm$ 0.2\\ 
&3D-CNN & 95.58 $\pm$ 0.3 & 97.03 $\pm$ 0.9 & 96.39 $\pm$ 0.2 \\ 
& SSRN & 99.87 $\pm$ 0.0 & 99.91 $\pm$ 0.0 & 99.90 $\pm$ 0.0\\
& HybridSN & 99.98 $\pm$ 0.0 & 99.97 $\pm$ 0.0 & 99.98 $\pm$ 0.0\\
& HSSNB & 99.98 $\pm$ 0.0 & 99.97 $\pm$ 0.0 & 99.98 $\pm$ 0.0\\
\hline

\hline
 SA  &2D-CNN  & 97.09 $\pm$ 0.1 & 98.81 $\pm$ 0.1  & 97.40 $\pm$ 0.0\\ 
&3D-CNN & 93.58 $\pm$ 0.5 & 97.05 $\pm$ 0.6 & 93.99 $\pm$ 0.2 \\ 
& SSRN & 99.97 $\pm$ 0.1 & 99.97 $\pm$ 0.0 & 99.98 $\pm$ 0.1\\
& HybridSN & 100 $\pm$ 0.0 & 100 $\pm$ 0.0 & 100 $\pm$ 0.0\\
& HSSNB & 100 $\pm$ 0.0 & 100 $\pm$ 0.0 & 100 $\pm$ 0.0\\
\hline

\hline
\end{tabular}
\end{table}

\begin{figure*}[!t]
\centering
\subfloat[]{\includegraphics[width=2in]{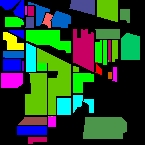}%
\label{Ground_truth_IP}}
\hfil
\subfloat[]{\includegraphics[width=2in]{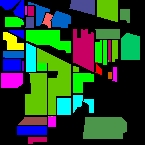}%
\label{Predicted_image_IP}}
\caption{ (a) IP data ground truth (b) IP Data predicted image.}
\label{fig_IP}
\end{figure*}

\begin{figure*}[!t]
\centering
\subfloat[]{\includegraphics[width=2in]{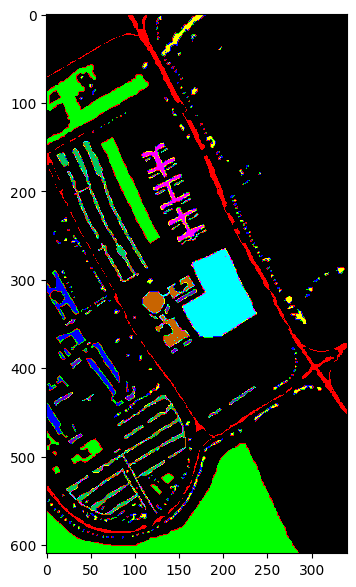}%
\label{Ground_truth_PU}}
\hfil
\subfloat[]{\includegraphics[width=2in]{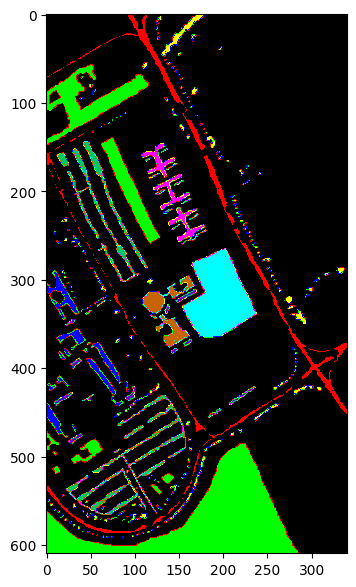}%
\label{Predicted_image_PU}}
\caption{ (a) PU data ground truth (b) PU Data predicted image.}
\label{fig_PU}
\end{figure*}

\begin{figure*}[!t]
\centering
\subfloat[]{\includegraphics[width=2in]{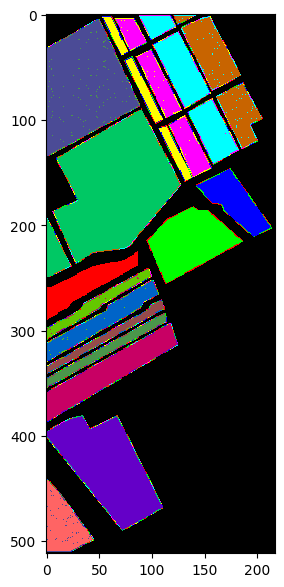}%
\label{Ground_truth_PU}}
\hfil
\subfloat[]{\includegraphics[width=2in]{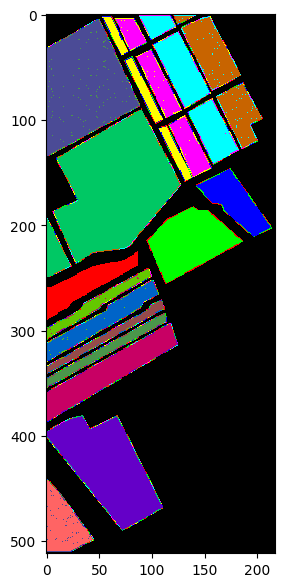}%
\label{Predicted_image_PU}}
\caption{ (a) SA data ground truth (b) SA Data predicted image.}
\label{fig_SA}
\end{figure*}

\begin{figure*}[!t]
\centering
\subfloat[]{\includegraphics[width=2in]{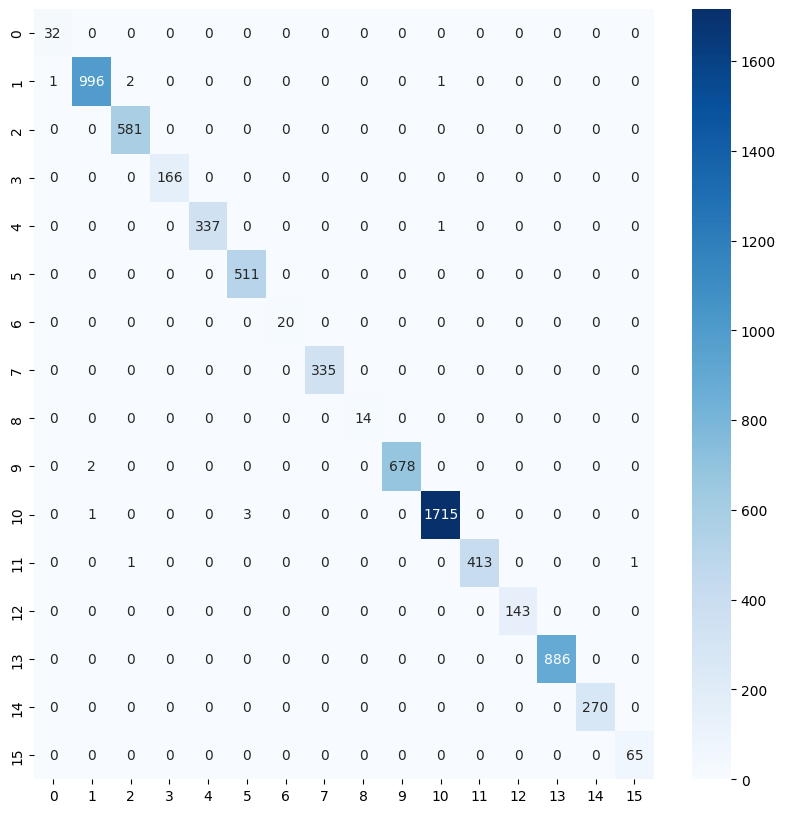}%
\label{Confusionmatrix for IP dataset}}
\hfil
\subfloat[]{\includegraphics[width=2in]{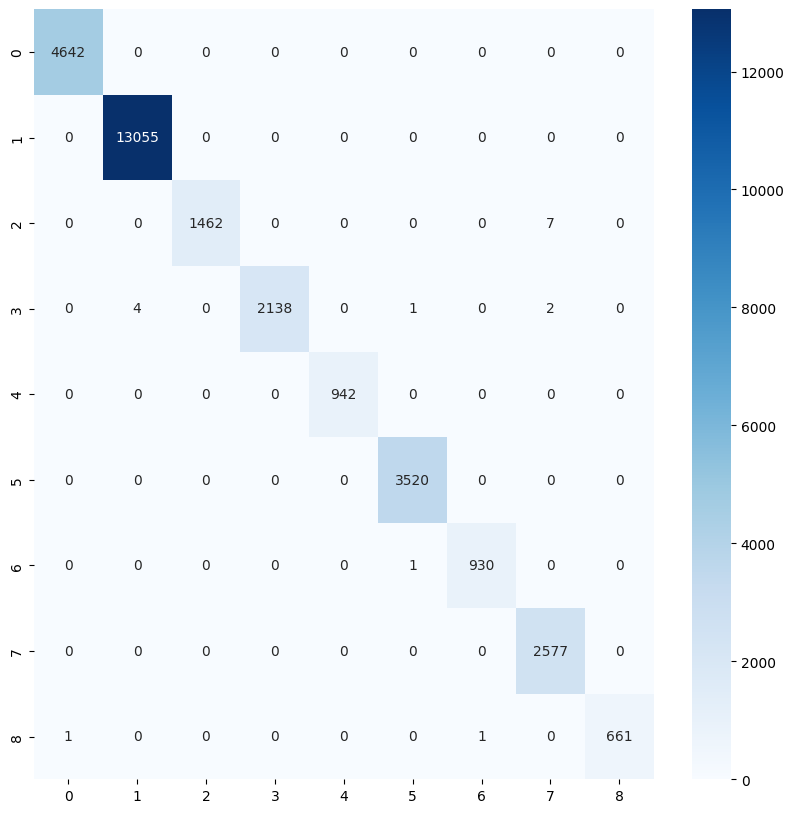}%
\label{Confusionmatrix for PU dataset}}
\hfil
\subfloat[]{\includegraphics[width=2in]{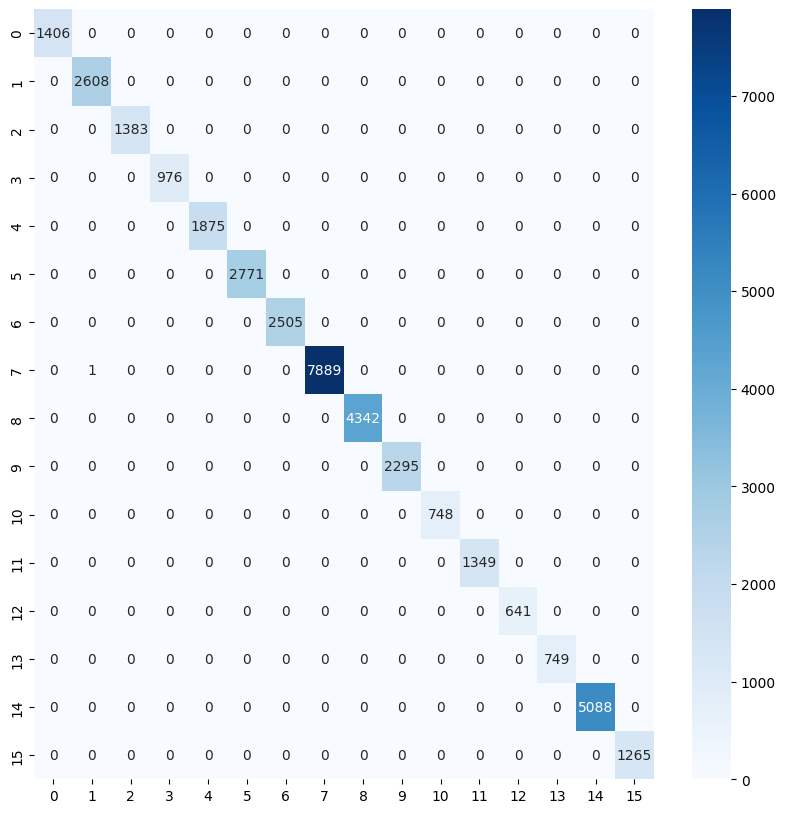}%
\hfil
\label{Confusionmatrix for SA dataset}}
\caption{Confusion matrix for  (a) IP data (b) PU Data (c) SA data}
\label{fig_confusionmatrix}
\end{figure*}

\begin{figure*}[!t]
\centering
\subfloat[]{\includegraphics[width=2in]{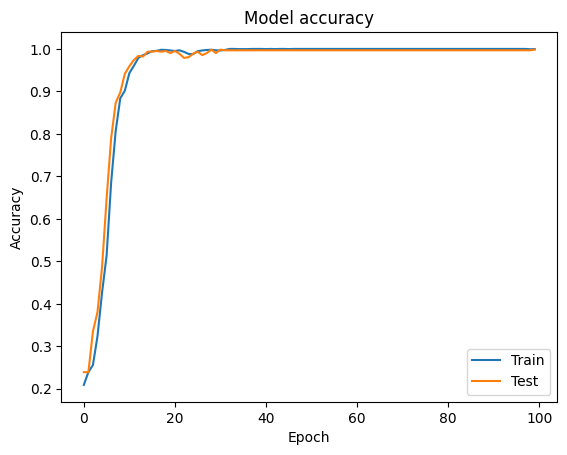}%
\label{Epoch vs Accuracy  for IP dataset}}
\hfil
\subfloat[]{\includegraphics[width=2in]{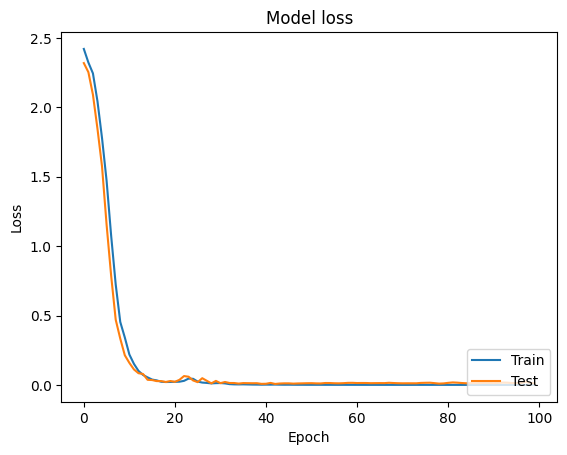}%
\label{Epoch vs Loss  for IP dataset}}
\
\caption{(a) Model accuracy and (b)model loss with respect to Epochs for IP dataset}
\label{fig_epoch_IP}
\end{figure*}

\begin{figure*}[!t]
\centering
\subfloat[]{\includegraphics[width=2in]{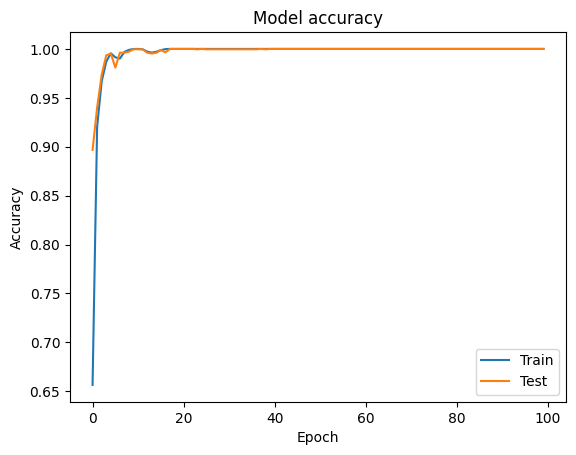}%
\label{Epoch vs Accuracy  for PU dataset}}
\hfil
\subfloat[]{\includegraphics[width=2in]{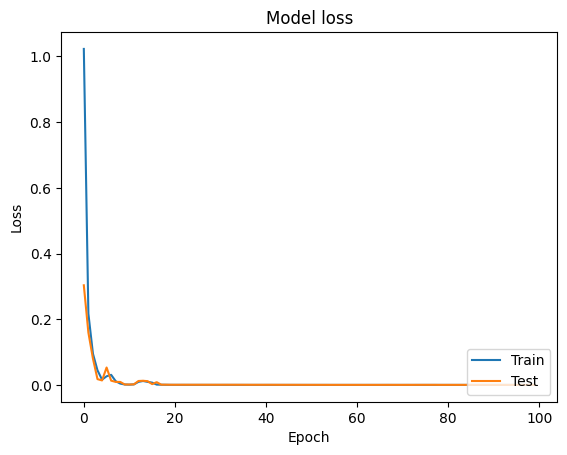}%
\label{Epoch vs Loss  for PU dataset}}
\
\caption{(a) Model accuracy and (b) model loss with respect to Epochs for PU dataset}
\label{fig_epoch_PU}
\end{figure*}

\begin{figure*}[!t]
\centering
\subfloat[]{\includegraphics[width=2in]{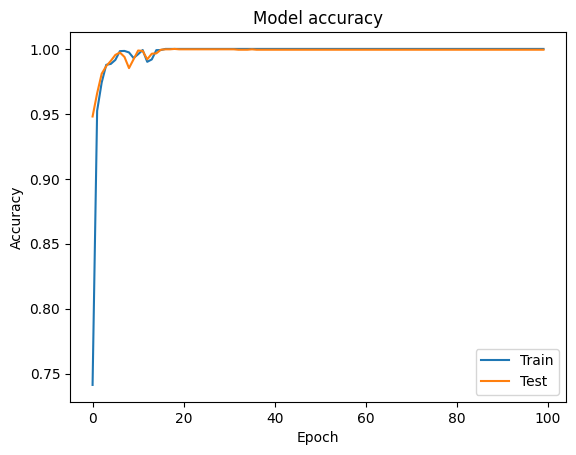}%
\label{Epoch vs Accuracy  for SA dataset}}
\hfil
\subfloat[]{\includegraphics[width=2in]{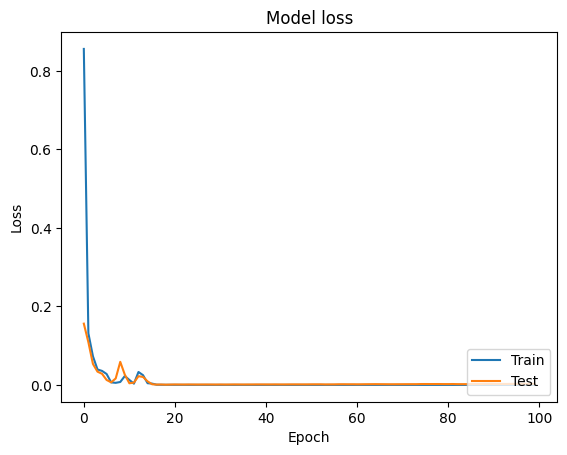}%
\label{Epoch vs Loss  for SA dataset}}
\
\caption{(a) Model accuracy and (b) model loss with respect to Epochs for SA dataset}
\label{fig_epoch_SA}
\end{figure*}

 \begin{table}[!t]
\caption{Accuracy with respect to window size \label{Windowsize}}
\centering
\begin{tabular}{|c||c||c||}

\hline
Dataset & Windowsize &    OA\\      
\hline
IP 
&19 $\times$ 19& 99.79\\
&21 $\times$ 21 &  99.29\\ 
&23  $\times$  23 &  99.74\\ 
&25  $\times$  25 &  99.83\\
\hline
PU
&19$\times$ 19 &  99.98\\
&21$\times$ 21  &  99.97\\ 
&23 $\times$ 23 &  99.91\\ 
&25 $\times$ 25 &  99.98\\
\hline
SA 
&19 $\times$ 19&  99.99\\
&21 $\times$ 21 &  99.75\\ 
&23 $\times$ 23 &  99.81\\ 
&25 $\times$ 25 &  100.00\\

\hline
\end{tabular}
\end{table}

We used the above discussed data sets (IP, PU, SA ) for evaluating the performance of our model. The training and testing ratio was 30:70. Kappa coefficient (Kappa),  average accuracy (AA) and overall accuracy (OA) were used as the performance measures of the model. Kappa coefficient was calculated using the matrix which shows the correlation between the actual ground truth map and the predicted map. Google Colab platform has been used for all of the experiments. The T4 GPU from Nvidia was selected as the runtype. We have also used NVIDIA GeForce GTX 1080 Ti GPU for comparisons with the state-of-art models.   

\subsection{Finalizing the window size}

To finalise the window size, we tested it for  4 different window sizes (19 $\times$ 19, 21 $\times$ 21, 23 $\times$ 23 and 25 $\times$ 25). It was tested on all the above datasets. From the table \ref{Windowsize}, we can observe that  25 $\times$ 25 window size is most suitable for these datasets. Hence, we did all the further experiments using the aforesaid window size. 

\subsection{Analysis of results}

We tested our model on IP, SA, PU datasets. The results of our model has been shown in the table \ref{comparativestudy}. The results of our model are compared with HybridSN\cite{roy2019hybridsn}, SSRN\cite{SSRN}, 3D-CNN\cite{3D-CNN} and 2D-CNN\cite{makantasis2015deep}. From the table \ref{comparativestudy}, we can observe that for IP dataset, HSSNB achieved $99.80 \pm 0.1$,  $99.89 \pm 0.1$ and $99.83 \pm 0.1$ in terms of kappa, AA and OA respectively. It was higher than HybridSN model by 0.09\%(kappa), 0.26\%(AA) and 0.08\%(OA). Fig \ref{fig_IP} shows the classification result for IP data. In case of PU dataset, HSSNB achieved the same accuracy as HybridSN.  Fig \ref{fig_PU} shows the classification result for it. For SA dataset, our model achieved 100\% accuracy. It is evident in fig  \ref{fig_SA}. 

The interesting part is that the model used only 15,43,040 trainable parameters. HybridSN model used 51,22,176 training parameters. 
The Hybrid model of 3D CNN and 2D CNN have been shown to be learning both spatial and spectral data effectively. It is evident from the results shown in the table \ref{comparativestudy}. However, this model is having information loss problem. It also can not learn inter-layer information \cite{convlstm}. Hence, the use of Bi-LSTM along with the 3D and 2D CNN stack solves that problem. Moreover, Bi-LSTMs can also learn correlations in long range \cite{vyawahare2022pict}. Hence, our model could perform better.   

So, it can be said that HSSNB achieved better accuracy than HybridSN by using only 30\% training parameters. Fig \ref{fig_confusionmatrix} shows the confusion matrices for IP, PU, SA datasets respectively. Fig \ref{fig_epoch_IP}, Fig \ref{fig_epoch_PU} and Fig \ref{fig_epoch_SA} shows the accuracy and training loss with respect to the number of epochs for all three datasets.

\begin{table}[!t]
\caption{Comparative study with 10 percent data \label{10percentdata}}
\centering
\begin{tabular}{|c||c||c||c||c|}

\hline
Dataset &Models &    Kappa & AA & OA\\      
\hline
IP  &2D-CNN  & 79.10  & 70.19  & 81.15 \\ 
&3D-CNN & 80.55 & 77.92  & 83.38  \\ 
& SSRN & 98.01& 90.92 & 98.15 \\
& HybridSN & 98.16  & 98.01 & 98.39\\
& HSSNB & 98.42  & 98.15  & 98.62 \\
\hline

\hline
 PU  &2D-CNN  & 95.16 & 95.01  & 97.01 \\ 
&3D-CNN & 95.59 & 96.89 & 96.39  \\ 
& SSRN & 99.47  & 99.51  & 99.60 \\
& HybridSN & 99.64  & 99.20  & 99.72 \\
& HSSNB & 99.60  & 99.28  & 99.64\\
\hline

\hline
 SA  &2D-CNN  & 95.09  & 95.81  & 96.40 \\ 
&3D-CNN & 86.38  & 89.50  & 87.59 \\ 
& SSRN & 99.67 & 99.79 & 99.68 \\
& HybridSN & 99.98 & 99.98 & 99.98 \\
& HSSNB & 99.98 & 99.98 & 99.98 \\
\hline

\hline
\end{tabular}
\end{table}

\subsection{Experimentation with 10\% training data}

We reduced the training samples to 10\% to evaluate the performance of the model with less training samples. Table \ref{10percentdata} shows the results for it. It can be observed that for IP dataset, HSSNB achieved 98.42\%, 98.15\%  and 98.62\% in terms of kappa, AA and OA respectively. It outperformed HybridSN by 0.26\%, 0.14\% and 0.21\%. For PU dataset, the accuracy of HybridSN is slightly higher than HSSNB. our model achieved same accuracy for SA dataset with 10\% training dataset as well.  

For IP dataset, the accuracy of HSSNB reduced by 1.38\%, 1.74\% and 1.21\% in terms of kappa, AA and OA respectively with respect to the model trained with 30\% training data. For PU and SA the difference is not significant. However, the model is able to outperform other models trained with 10\% training dataset. So, it can be said that the model can perform well with reduced training data.

\section{Conclusion}
This paper proposed a model named HSSNB combining 3-D CNN, 2-D CNN and Bi-LSTM. The main target of this paper was to develop a model that uses less number of training parameters while giving better accuracy. The proposed model was tested on 3 datasets (IP, SA, PU ). The model outperformed the state of art models by 0.09\%, 0.26\% and 0.08\% in terms of kappa, AA and OA respectively for IP dataset. However, for PU and SA, the accuracy remained same. The main advantage of this model is that it achieved better results using only 30\% training parameters of the state-of-art model. Hence, it can be safely said that HSSNB could perform better using significantly less number of parameters (70\% less). It could also perform better than other state of art models using only 10 percent train data. We further want to decrease the training parameters of the model in our further studies.

\section*{Statements and Declarations}

\begin{itemize}
\item The authors have no relevant financial or non-financial interests to disclose.
\item The authors have no competing interests to declare that are relevant to the content of this article.
\item All authors certify that they have no affiliations with or involvement in any organization or entity with any financial interest or non-financial interest in the subject matter or materials discussed in this manuscript.
\item The authors have no financial or proprietary interests in any material discussed in this article.
\item Availability of data and materials: Not Applicable
\item Code availability : Not Applicable
\end{itemize}

\bibliography{sn-bibliography}


\begin{thebibliography}{44}
\ifx \bisbn   \undefined \def \bisbn  #1{ISBN #1}\fi
\ifx \binits  \undefined \def \binits#1{#1}\fi
\ifx \bauthor  \undefined \def \bauthor#1{#1}\fi
\ifx \batitle  \undefined \def \batitle#1{#1}\fi
\ifx \bjtitle  \undefined \def \bjtitle#1{#1}\fi
\ifx \bvolume  \undefined \def \bvolume#1{\textbf{#1}}\fi
\ifx \byear  \undefined \def \byear#1{#1}\fi
\ifx \bissue  \undefined \def \bissue#1{#1}\fi
\ifx \bfpage  \undefined \def \bfpage#1{#1}\fi
\ifx \blpage  \undefined \def \blpage #1{#1}\fi
\ifx \burl  \undefined \def \burl#1{\textsf{#1}}\fi
\ifx \doiurl  \undefined \def \doiurl#1{\url{https://doi.org/#1}}\fi
\ifx \betal  \undefined \def \betal{\textit{et al.}}\fi
\ifx \binstitute  \undefined \def \binstitute#1{#1}\fi
\ifx \binstitutionaled  \undefined \def \binstitutionaled#1{#1}\fi
\ifx \bctitle  \undefined \def \bctitle#1{#1}\fi
\ifx \beditor  \undefined \def \beditor#1{#1}\fi
\ifx \bpublisher  \undefined \def \bpublisher#1{#1}\fi
\ifx \bbtitle  \undefined \def \bbtitle#1{#1}\fi
\ifx \bedition  \undefined \def \bedition#1{#1}\fi
\ifx \bseriesno  \undefined \def \bseriesno#1{#1}\fi
\ifx \blocation  \undefined \def \blocation#1{#1}\fi
\ifx \bsertitle  \undefined \def \bsertitle#1{#1}\fi
\ifx \bsnm \undefined \def \bsnm#1{#1}\fi
\ifx \bsuffix \undefined \def \bsuffix#1{#1}\fi
\ifx \bparticle \undefined \def \bparticle#1{#1}\fi
\ifx \barticle \undefined \def \barticle#1{#1}\fi
\bibcommenthead
\ifx \bconfdate \undefined \def \bconfdate #1{#1}\fi
\ifx \botherref \undefined \def \botherref #1{#1}\fi
\ifx \url \undefined \def \url#1{\textsf{#1}}\fi
\ifx \bchapter \undefined \def \bchapter#1{#1}\fi
\ifx \bbook \undefined \def \bbook#1{#1}\fi
\ifx \bcomment \undefined \def \bcomment#1{#1}\fi
\ifx \oauthor \undefined \def \oauthor#1{#1}\fi
\ifx \citeauthoryear \undefined \def \citeauthoryear#1{#1}\fi
\ifx \endbibitem  \undefined \def \endbibitem {}\fi
\ifx \bconflocation  \undefined \def \bconflocation#1{#1}\fi
\ifx \arxivurl  \undefined \def \arxivurl#1{\textsf{#1}}\fi
\csname PreBibitemsHook\endcsname

\bibitem[\protect\citeauthoryear{Gevaert et~al.}{2015}]{gevaert2015generation}
\begin{barticle}
\bauthor{\bsnm{Gevaert}, \binits{C.M.}},
\bauthor{\bsnm{Suomalainen}, \binits{J.}},
\bauthor{\bsnm{Tang}, \binits{J.}},
\bauthor{\bsnm{Kooistra}, \binits{L.}}:
\batitle{Generation of spectral--temporal response surfaces by combining multispectral satellite and hyperspectral uav imagery for precision agriculture applications}.
\bjtitle{IEEE Journal of Selected Topics in Applied Earth Observations and Remote Sensing}
\bvolume{8}(\bissue{6}),
\bfpage{3140}--\blpage{3146}
(\byear{2015})
\end{barticle}
\endbibitem

\bibitem[\protect\citeauthoryear{Shimoni et~al.}{2019}]{shimoni2019hypersectral}
\begin{barticle}
\bauthor{\bsnm{Shimoni}, \binits{M.}},
\bauthor{\bsnm{Haelterman}, \binits{R.}},
\bauthor{\bsnm{Perneel}, \binits{C.}}:
\batitle{Hypersectral imaging for military and security applications: Combining myriad processing and sensing techniques}.
\bjtitle{IEEE Geoscience and Remote Sensing Magazine}
\bvolume{7}(\bissue{2}),
\bfpage{101}--\blpage{117}
(\byear{2019})
\end{barticle}
\endbibitem

\bibitem[\protect\citeauthoryear{Sudharsan et~al.}{2019}]{sudharsan2019survey}
\begin{bchapter}
\bauthor{\bsnm{Sudharsan}, \binits{S.}},
\bauthor{\bsnm{Hemalatha}, \binits{R.}},
\bauthor{\bsnm{Radha}, \binits{S.}}:
\bctitle{A survey on hyperspectral imaging for mineral exploration using machine learning algorithms}.
In: \bbtitle{2019 International Conference on Wireless Communications Signal Processing and Networking (WiSPNET)},
pp. \bfpage{206}--\blpage{212}
(\byear{2019}).
\bcomment{IEEE}
\end{bchapter}
\endbibitem

\bibitem[\protect\citeauthoryear{Ortac and Ozcan}{2021}]{ortac2021comparative}
\begin{barticle}
\bauthor{\bsnm{Ortac}, \binits{G.}},
\bauthor{\bsnm{Ozcan}, \binits{G.}}:
\batitle{Comparative study of hyperspectral image classification by multidimensional convolutional neural network approaches to improve accuracy}.
\bjtitle{Expert Systems with Applications}
\bvolume{182},
\bfpage{115280}
(\byear{2021})
\end{barticle}
\endbibitem

\bibitem[\protect\citeauthoryear{ul~Rehman and Qureshi}{2021}]{ul2021review}
\begin{barticle}
\bauthor{\bsnm{Rehman}, \binits{A.}},
\bauthor{\bsnm{Qureshi}, \binits{S.A.}}:
\batitle{A review of the medical hyperspectral imaging systems and unmixing algorithms’ in biological tissues}.
\bjtitle{Photodiagnosis and Photodynamic Therapy}
\bvolume{33},
\bfpage{102165}
(\byear{2021})
\end{barticle}
\endbibitem

\bibitem[\protect\citeauthoryear{Lu and Fei}{2014}]{lu2014medical}
\begin{barticle}
\bauthor{\bsnm{Lu}, \binits{G.}},
\bauthor{\bsnm{Fei}, \binits{B.}}:
\batitle{Medical hyperspectral imaging: a review}.
\bjtitle{Journal of biomedical optics}
\bvolume{19}(\bissue{1}),
\bfpage{010901}--\blpage{010901}
(\byear{2014})
\end{barticle}
\endbibitem

\bibitem[\protect\citeauthoryear{Scholkopf and Smola}{2018}]{scholkopf2018learning}
\begin{bbook}
\bauthor{\bsnm{Scholkopf}, \binits{B.}},
\bauthor{\bsnm{Smola}, \binits{A.J.}}:
\bbtitle{Learning with Kernels: Support Vector Machines, Regularization, Optimization, and Beyond}.
\bpublisher{MIT press}, \blocation{???}
(\byear{2018})
\end{bbook}
\endbibitem

\bibitem[\protect\citeauthoryear{Bo et~al.}{2015}]{bo2015hyperspectral}
\begin{barticle}
\bauthor{\bsnm{Bo}, \binits{C.}},
\bauthor{\bsnm{Lu}, \binits{H.}},
\bauthor{\bsnm{Wang}, \binits{D.}}:
\batitle{Hyperspectral image classification via jcr and svm models with decision fusion}.
\bjtitle{IEEE Geoscience and Remote Sensing Letters}
\bvolume{13}(\bissue{2}),
\bfpage{177}--\blpage{181}
(\byear{2015})
\end{barticle}
\endbibitem

\bibitem[\protect\citeauthoryear{Li et~al.}{2010}]{li2010semisupervised}
\begin{barticle}
\bauthor{\bsnm{Li}, \binits{J.}},
\bauthor{\bsnm{Bioucas-Dias}, \binits{J.M.}},
\bauthor{\bsnm{Plaza}, \binits{A.}}:
\batitle{Semisupervised hyperspectral image segmentation using multinomial logistic regression with active learning}.
\bjtitle{IEEE Transactions on Geoscience and Remote Sensing}
\bvolume{48}(\bissue{11}),
\bfpage{4085}--\blpage{4098}
(\byear{2010})
\end{barticle}
\endbibitem

\bibitem[\protect\citeauthoryear{Li et~al.}{2011}]{li2011spectral}
\begin{barticle}
\bauthor{\bsnm{Li}, \binits{J.}},
\bauthor{\bsnm{Bioucas-Dias}, \binits{J.M.}},
\bauthor{\bsnm{Plaza}, \binits{A.}}:
\batitle{Spectral--spatial hyperspectral image segmentation using subspace multinomial logistic regression and markov random fields}.
\bjtitle{IEEE Transactions on Geoscience and Remote Sensing}
\bvolume{50}(\bissue{3}),
\bfpage{809}--\blpage{823}
(\byear{2011})
\end{barticle}
\endbibitem

\bibitem[\protect\citeauthoryear{Du and Zhang}{2010}]{du2010random}
\begin{barticle}
\bauthor{\bsnm{Du}, \binits{B.}},
\bauthor{\bsnm{Zhang}, \binits{L.}}:
\batitle{Random-selection-based anomaly detector for hyperspectral imagery}.
\bjtitle{IEEE Transactions on Geoscience and Remote sensing}
\bvolume{49}(\bissue{5}),
\bfpage{1578}--\blpage{1589}
(\byear{2010})
\end{barticle}
\endbibitem

\bibitem[\protect\citeauthoryear{Du and Zhang}{2014}]{du2014target}
\begin{barticle}
\bauthor{\bsnm{Du}, \binits{B.}},
\bauthor{\bsnm{Zhang}, \binits{L.}}:
\batitle{Target detection based on a dynamic subspace}.
\bjtitle{Pattern Recognition}
\bvolume{47}(\bissue{1}),
\bfpage{344}--\blpage{358}
(\byear{2014})
\end{barticle}
\endbibitem

\bibitem[\protect\citeauthoryear{Bandos et~al.}{2009}]{bandos2009classification}
\begin{barticle}
\bauthor{\bsnm{Bandos}, \binits{T.V.}},
\bauthor{\bsnm{Bruzzone}, \binits{L.}},
\bauthor{\bsnm{Camps-Valls}, \binits{G.}}:
\batitle{Classification of hyperspectral images with regularized linear discriminant analysis}.
\bjtitle{IEEE Transactions on Geoscience and Remote Sensing}
\bvolume{47}(\bissue{3}),
\bfpage{862}--\blpage{873}
(\byear{2009})
\end{barticle}
\endbibitem

\bibitem[\protect\citeauthoryear{Prasad and Bruce}{2008}]{prasad2008limitations}
\begin{barticle}
\bauthor{\bsnm{Prasad}, \binits{S.}},
\bauthor{\bsnm{Bruce}, \binits{L.M.}}:
\batitle{Limitations of principal components analysis for hyperspectral target recognition}.
\bjtitle{IEEE Geoscience and Remote Sensing Letters}
\bvolume{5}(\bissue{4}),
\bfpage{625}--\blpage{629}
(\byear{2008})
\end{barticle}
\endbibitem

\bibitem[\protect\citeauthoryear{Licciardi et~al.}{2011}]{licciardi2011linear}
\begin{barticle}
\bauthor{\bsnm{Licciardi}, \binits{G.}},
\bauthor{\bsnm{Marpu}, \binits{P.R.}},
\bauthor{\bsnm{Chanussot}, \binits{J.}},
\bauthor{\bsnm{Benediktsson}, \binits{J.A.}}:
\batitle{Linear versus nonlinear pca for the classification of hyperspectral data based on the extended morphological profiles}.
\bjtitle{IEEE Geoscience and Remote Sensing Letters}
\bvolume{9}(\bissue{3}),
\bfpage{447}--\blpage{451}
(\byear{2011})
\end{barticle}
\endbibitem

\bibitem[\protect\citeauthoryear{Villa et~al.}{2011}]{villa2011hyperspectral}
\begin{barticle}
\bauthor{\bsnm{Villa}, \binits{A.}},
\bauthor{\bsnm{Benediktsson}, \binits{J.A.}},
\bauthor{\bsnm{Chanussot}, \binits{J.}},
\bauthor{\bsnm{Jutten}, \binits{C.}}:
\batitle{Hyperspectral image classification with independent component discriminant analysis}.
\bjtitle{IEEE transactions on Geoscience and remote sensing}
\bvolume{49}(\bissue{12}),
\bfpage{4865}--\blpage{4876}
(\byear{2011})
\end{barticle}
\endbibitem

\bibitem[\protect\citeauthoryear{Krishnapuram et~al.}{2005}]{krishnapuram2005sparse}
\begin{barticle}
\bauthor{\bsnm{Krishnapuram}, \binits{B.}},
\bauthor{\bsnm{Carin}, \binits{L.}},
\bauthor{\bsnm{Figueiredo}, \binits{M.A.}},
\bauthor{\bsnm{Hartemink}, \binits{A.J.}}:
\batitle{Sparse multinomial logistic regression: Fast algorithms and generalization bounds}.
\bjtitle{IEEE transactions on pattern analysis and machine intelligence}
\bvolume{27}(\bissue{6}),
\bfpage{957}--\blpage{968}
(\byear{2005})
\end{barticle}
\endbibitem

\bibitem[\protect\citeauthoryear{Wang et~al.}{2017}]{wang2017locality}
\begin{barticle}
\bauthor{\bsnm{Wang}, \binits{Q.}},
\bauthor{\bsnm{Meng}, \binits{Z.}},
\bauthor{\bsnm{Li}, \binits{X.}}:
\batitle{Locality adaptive discriminant analysis for spectral--spatial classification of hyperspectral images}.
\bjtitle{IEEE Geoscience and Remote Sensing Letters}
\bvolume{14}(\bissue{11}),
\bfpage{2077}--\blpage{2081}
(\byear{2017})
\end{barticle}
\endbibitem

\bibitem[\protect\citeauthoryear{Fang et~al.}{2017}]{fang2017hyperspectral}
\begin{barticle}
\bauthor{\bsnm{Fang}, \binits{L.}},
\bauthor{\bsnm{Wang}, \binits{C.}},
\bauthor{\bsnm{Li}, \binits{S.}},
\bauthor{\bsnm{Benediktsson}, \binits{J.A.}}:
\batitle{Hyperspectral image classification via multiple-feature-based adaptive sparse representation}.
\bjtitle{IEEE Transactions on Instrumentation and Measurement}
\bvolume{66}(\bissue{7}),
\bfpage{1646}--\blpage{1657}
(\byear{2017})
\end{barticle}
\endbibitem

\bibitem[\protect\citeauthoryear{Guo et~al.}{2019}]{guo2019hyperspectral}
\begin{barticle}
\bauthor{\bsnm{Guo}, \binits{Y.}},
\bauthor{\bsnm{Yin}, \binits{X.}},
\bauthor{\bsnm{Zhao}, \binits{X.}},
\bauthor{\bsnm{Yang}, \binits{D.}},
\bauthor{\bsnm{Bai}, \binits{Y.}}:
\batitle{Hyperspectral image classification with svm and guided filter}.
\bjtitle{EURASIP Journal on Wireless Communications and Networking}
\bvolume{2019}(\bissue{1}),
\bfpage{1}--\blpage{9}
(\byear{2019})
\end{barticle}
\endbibitem

\bibitem[\protect\citeauthoryear{Banerjee and Banik}{2023}]{banerjee2023pooled}
\begin{barticle}
\bauthor{\bsnm{Banerjee}, \binits{A.}},
\bauthor{\bsnm{Banik}, \binits{D.}}:
\batitle{Pooled hybrid-spectral for hyperspectral image classification}.
\bjtitle{Multimedia Tools and Applications}
\bvolume{82}(\bissue{7}),
\bfpage{10887}--\blpage{10899}
(\byear{2023})
\end{barticle}
\endbibitem

\bibitem[\protect\citeauthoryear{Xu et~al.}{2019}]{xu2019d}
\begin{barticle}
\bauthor{\bsnm{Xu}, \binits{M.}},
\bauthor{\bsnm{Fang}, \binits{H.}},
\bauthor{\bsnm{Lv}, \binits{P.}},
\bauthor{\bsnm{Cui}, \binits{L.}},
\bauthor{\bsnm{Zhang}, \binits{S.}},
\bauthor{\bsnm{Zhou}, \binits{B.}}:
\batitle{D-stc: Deep learning with spatio-temporal constraints for train drivers detection from videos}.
\bjtitle{Pattern Recognition Letters}
\bvolume{119},
\bfpage{222}--\blpage{228}
(\byear{2019})
\end{barticle}
\endbibitem

\bibitem[\protect\citeauthoryear{Kang et~al.}{2018}]{kang2018dual}
\begin{barticle}
\bauthor{\bsnm{Kang}, \binits{X.}},
\bauthor{\bsnm{Zhuo}, \binits{B.}},
\bauthor{\bsnm{Duan}, \binits{P.}}:
\batitle{Dual-path network-based hyperspectral image classification}.
\bjtitle{IEEE Geoscience and Remote Sensing Letters}
\bvolume{16}(\bissue{3}),
\bfpage{447}--\blpage{451}
(\byear{2018})
\end{barticle}
\endbibitem

\bibitem[\protect\citeauthoryear{Yu et~al.}{2017}]{yu2017unsupervised}
\begin{barticle}
\bauthor{\bsnm{Yu}, \binits{Y.}},
\bauthor{\bsnm{Gong}, \binits{Z.}},
\bauthor{\bsnm{Wang}, \binits{C.}},
\bauthor{\bsnm{Zhong}, \binits{P.}}:
\batitle{An unsupervised convolutional feature fusion network for deep representation of remote sensing images}.
\bjtitle{IEEE Geoscience and Remote Sensing Letters}
\bvolume{15}(\bissue{1}),
\bfpage{23}--\blpage{27}
(\byear{2017})
\end{barticle}
\endbibitem

\bibitem[\protect\citeauthoryear{Song et~al.}{2018}]{song2018hyperspectral}
\begin{barticle}
\bauthor{\bsnm{Song}, \binits{W.}},
\bauthor{\bsnm{Li}, \binits{S.}},
\bauthor{\bsnm{Fang}, \binits{L.}},
\bauthor{\bsnm{Lu}, \binits{T.}}:
\batitle{Hyperspectral image classification with deep feature fusion network}.
\bjtitle{IEEE Transactions on Geoscience and Remote Sensing}
\bvolume{56}(\bissue{6}),
\bfpage{3173}--\blpage{3184}
(\byear{2018})
\end{barticle}
\endbibitem

\bibitem[\protect\citeauthoryear{Cheng et~al.}{2018}]{cheng2018exploring}
\begin{barticle}
\bauthor{\bsnm{Cheng}, \binits{G.}},
\bauthor{\bsnm{Li}, \binits{Z.}},
\bauthor{\bsnm{Han}, \binits{J.}},
\bauthor{\bsnm{Yao}, \binits{X.}},
\bauthor{\bsnm{Guo}, \binits{L.}}:
\batitle{Exploring hierarchical convolutional features for hyperspectral image classification}.
\bjtitle{IEEE Transactions on Geoscience and Remote Sensing}
\bvolume{56}(\bissue{11}),
\bfpage{6712}--\blpage{6722}
(\byear{2018})
\end{barticle}
\endbibitem

\bibitem[\protect\citeauthoryear{Chen et~al.}{2016}]{chen2016deep}
\begin{barticle}
\bauthor{\bsnm{Chen}, \binits{Y.}},
\bauthor{\bsnm{Jiang}, \binits{H.}},
\bauthor{\bsnm{Li}, \binits{C.}},
\bauthor{\bsnm{Jia}, \binits{X.}},
\bauthor{\bsnm{Ghamisi}, \binits{P.}}:
\batitle{Deep feature extraction and classification of hyperspectral images based on convolutional neural networks}.
\bjtitle{IEEE transactions on geoscience and remote sensing}
\bvolume{54}(\bissue{10}),
\bfpage{6232}--\blpage{6251}
(\byear{2016})
\end{barticle}
\endbibitem

\bibitem[\protect\citeauthoryear{Zhong et~al.}{2017}]{zhong2017spectral}
\begin{barticle}
\bauthor{\bsnm{Zhong}, \binits{Z.}},
\bauthor{\bsnm{Li}, \binits{J.}},
\bauthor{\bsnm{Luo}, \binits{Z.}},
\bauthor{\bsnm{Chapman}, \binits{M.}}:
\batitle{Spectral--spatial residual network for hyperspectral image classification: A 3-d deep learning framework}.
\bjtitle{IEEE Transactions on Geoscience and Remote Sensing}
\bvolume{56}(\bissue{2}),
\bfpage{847}--\blpage{858}
(\byear{2017})
\end{barticle}
\endbibitem

\bibitem[\protect\citeauthoryear{Mou et~al.}{2017}]{mou2017unsupervised}
\begin{barticle}
\bauthor{\bsnm{Mou}, \binits{L.}},
\bauthor{\bsnm{Ghamisi}, \binits{P.}},
\bauthor{\bsnm{Zhu}, \binits{X.X.}}:
\batitle{Unsupervised spectral--spatial feature learning via deep residual conv--deconv network for hyperspectral image classification}.
\bjtitle{IEEE Transactions on Geoscience and Remote Sensing}
\bvolume{56}(\bissue{1}),
\bfpage{391}--\blpage{406}
(\byear{2017})
\end{barticle}
\endbibitem

\bibitem[\protect\citeauthoryear{Paoletti et~al.}{2018}]{paoletti2018deep}
\begin{barticle}
\bauthor{\bsnm{Paoletti}, \binits{M.E.}},
\bauthor{\bsnm{Haut}, \binits{J.M.}},
\bauthor{\bsnm{Fernandez-Beltran}, \binits{R.}},
\bauthor{\bsnm{Plaza}, \binits{J.}},
\bauthor{\bsnm{Plaza}, \binits{A.J.}},
\bauthor{\bsnm{Pla}, \binits{F.}}:
\batitle{Deep pyramidal residual networks for spectral--spatial hyperspectral image classification}.
\bjtitle{IEEE Transactions on Geoscience and Remote Sensing}
\bvolume{57}(\bissue{2}),
\bfpage{740}--\blpage{754}
(\byear{2018})
\end{barticle}
\endbibitem

\bibitem[\protect\citeauthoryear{Roy et~al.}{2019}]{roy2019hybridsn}
\begin{barticle}
\bauthor{\bsnm{Roy}, \binits{S.K.}},
\bauthor{\bsnm{Krishna}, \binits{G.}},
\bauthor{\bsnm{Dubey}, \binits{S.R.}},
\bauthor{\bsnm{Chaudhuri}, \binits{B.B.}}:
\batitle{Hybridsn: Exploring 3-d--2-d cnn feature hierarchy for hyperspectral image classification}.
\bjtitle{IEEE Geoscience and Remote Sensing Letters}
\bvolume{17}(\bissue{2}),
\bfpage{277}--\blpage{281}
(\byear{2019})
\end{barticle}
\endbibitem

\bibitem[\protect\citeauthoryear{Hu et~al.}{2020}]{convlstm}
\begin{barticle}
\bauthor{\bsnm{Hu}, \binits{W.-S.}},
\bauthor{\bsnm{Li}, \binits{H.-C.}},
\bauthor{\bsnm{Pan}, \binits{L.}},
\bauthor{\bsnm{Li}, \binits{W.}},
\bauthor{\bsnm{Tao}, \binits{R.}},
\bauthor{\bsnm{Du}, \binits{Q.}}:
\batitle{Spatial--spectral feature extraction via deep convlstm neural networks for hyperspectral image classification}.
\bjtitle{IEEE Transactions on Geoscience and Remote Sensing}
\bvolume{58}(\bissue{6}),
\bfpage{4237}--\blpage{4250}
(\byear{2020})
\end{barticle}
\endbibitem

\bibitem[\protect\citeauthoryear{Zhou et~al.}{2019}]{ZHOU201939}
\begin{barticle}
\bauthor{\bsnm{Zhou}, \binits{F.}},
\bauthor{\bsnm{Hang}, \binits{R.}},
\bauthor{\bsnm{Liu}, \binits{Q.}},
\bauthor{\bsnm{Yuan}, \binits{X.}}:
\batitle{Hyperspectral image classification using spectral-spatial lstms}.
\bjtitle{Neurocomputing}
\bvolume{328},
\bfpage{39}--\blpage{47}
(\byear{2019})
\doiurl{10.1016/j.neucom.2018.02.105} .
\bcomment{Chinese Conference on Computer Vision 2017}
\end{barticle}
\endbibitem

\bibitem[\protect\citeauthoryear{Liu et~al.}{2017}]{liu2017bidirectional}
\begin{barticle}
\bauthor{\bsnm{Liu}, \binits{Q.}},
\bauthor{\bsnm{Zhou}, \binits{F.}},
\bauthor{\bsnm{Hang}, \binits{R.}},
\bauthor{\bsnm{Yuan}, \binits{X.}}:
\batitle{Bidirectional-convolutional lstm based spectral-spatial feature learning for hyperspectral image classification}.
\bjtitle{Remote Sensing}
\bvolume{9}(\bissue{12}),
\bfpage{1330}
(\byear{2017})
\end{barticle}
\endbibitem

\bibitem[\protect\citeauthoryear{Xu et~al.}{2022}]{9900270}
\begin{barticle}
\bauthor{\bsnm{Xu}, \binits{Q.}},
\bauthor{\bsnm{Yang}, \binits{C.}},
\bauthor{\bsnm{Tang}, \binits{J.}},
\bauthor{\bsnm{Luo}, \binits{B.}}:
\batitle{Grouped bidirectional lstm network and multistage fusion convolutional transformer for hyperspectral image classification}.
\bjtitle{IEEE Transactions on Geoscience and Remote Sensing}
\bvolume{60},
\bfpage{1}--\blpage{14}
(\byear{2022})
\doiurl{10.1109/TGRS.2022.3207294}
\end{barticle}
\endbibitem

\bibitem[\protect\citeauthoryear{Zhang et~al.}{2022}]{zhang2022caevt}
\begin{barticle}
\bauthor{\bsnm{Zhang}, \binits{Z.}},
\bauthor{\bsnm{Li}, \binits{T.}},
\bauthor{\bsnm{Tang}, \binits{X.}},
\bauthor{\bsnm{Hu}, \binits{X.}},
\bauthor{\bsnm{Peng}, \binits{Y.}}:
\batitle{Caevt: Convolutional autoencoder meets lightweight vision transformer for hyperspectral image classification}.
\bjtitle{Sensors}
\bvolume{22}(\bissue{10}),
\bfpage{3902}
(\byear{2022})
\end{barticle}
\endbibitem

\bibitem[\protect\citeauthoryear{Vyawahare et~al.}{2022}]{vyawahare2022pict}
\begin{botherref}
\oauthor{\bsnm{Vyawahare}, \binits{A.}},
\oauthor{\bsnm{Tangsali}, \binits{R.}},
\oauthor{\bsnm{Mandke}, \binits{A.}},
\oauthor{\bsnm{Litake}, \binits{O.}},
\oauthor{\bsnm{Kadam}, \binits{D.}}:
Pict@ dravidianlangtech-acl2022: Neural machine translation on dravidian languages.
arXiv preprint arXiv:2204.09098
(2022)
\end{botherref}
\endbibitem

\bibitem[\protect\citeauthoryear{Hochreiter et~al.}{2001}]{vanishinggradient}
\begin{botherref}
\oauthor{\bsnm{Hochreiter}, \binits{S.}},
\oauthor{\bsnm{Bengio}, \binits{Y.}},
\oauthor{\bsnm{Frasconi}, \binits{P.}},
\oauthor{\bsnm{Schmidhuber}, \binits{J.}}, et al.:
Gradient flow in recurrent nets: the difficulty of learning long-term dependencies.
A field guide to dynamical recurrent neural networks. IEEE Press In
(2001)
\end{botherref}
\endbibitem

\bibitem[\protect\citeauthoryear{Greff et~al.}{2016}]{vanilalstm}
\begin{barticle}
\bauthor{\bsnm{Greff}, \binits{K.}},
\bauthor{\bsnm{Srivastava}, \binits{R.K.}},
\bauthor{\bsnm{Koutn{\'\i}k}, \binits{J.}},
\bauthor{\bsnm{Steunebrink}, \binits{B.R.}},
\bauthor{\bsnm{Schmidhuber}, \binits{J.}}:
\batitle{Lstm: A search space odyssey}.
\bjtitle{IEEE transactions on neural networks and learning systems}
\bvolume{28}(\bissue{10}),
\bfpage{2222}--\blpage{2232}
(\byear{2016})
\end{barticle}
\endbibitem

\bibitem[\protect\citeauthoryear{Van~Houdt et~al.}{2020}]{LSTMreview}
\begin{barticle}
\bauthor{\bsnm{Van~Houdt}, \binits{G.}},
\bauthor{\bsnm{Mosquera}, \binits{C.}},
\bauthor{\bsnm{N{\'a}poles}, \binits{G.}}:
\batitle{A review on the long short-term memory model}.
\bjtitle{Artificial Intelligence Review}
\bvolume{53}(\bissue{8}),
\bfpage{5929}--\blpage{5955}
(\byear{2020})
\end{barticle}
\endbibitem

\bibitem[\protect\citeauthoryear{Ji et~al.}{2012}]{3dconv}
\begin{barticle}
\bauthor{\bsnm{Ji}, \binits{S.}},
\bauthor{\bsnm{Xu}, \binits{W.}},
\bauthor{\bsnm{Yang}, \binits{M.}},
\bauthor{\bsnm{Yu}, \binits{K.}}:
\batitle{3d convolutional neural networks for human action recognition}.
\bjtitle{IEEE transactions on pattern analysis and machine intelligence}
\bvolume{35}(\bissue{1}),
\bfpage{221}--\blpage{231}
(\byear{2012})
\end{barticle}
\endbibitem

\bibitem[\protect\citeauthoryear{Zhong et~al.}{2017}]{SSRN}
\begin{barticle}
\bauthor{\bsnm{Zhong}, \binits{Z.}},
\bauthor{\bsnm{Li}, \binits{J.}},
\bauthor{\bsnm{Luo}, \binits{Z.}},
\bauthor{\bsnm{Chapman}, \binits{M.}}:
\batitle{Spectral--spatial residual network for hyperspectral image classification: A 3-d deep learning framework}.
\bjtitle{IEEE Transactions on Geoscience and Remote Sensing}
\bvolume{56}(\bissue{2}),
\bfpage{847}--\blpage{858}
(\byear{2017})
\end{barticle}
\endbibitem

\bibitem[\protect\citeauthoryear{Hamida et~al.}{2018}]{3D-CNN}
\begin{barticle}
\bauthor{\bsnm{Hamida}, \binits{A.B.}},
\bauthor{\bsnm{Benoit}, \binits{A.}},
\bauthor{\bsnm{Lambert}, \binits{P.}},
\bauthor{\bsnm{Amar}, \binits{C.B.}}:
\batitle{3-d deep learning approach for remote sensing image classification}.
\bjtitle{IEEE Transactions on geoscience and remote sensing}
\bvolume{56}(\bissue{8}),
\bfpage{4420}--\blpage{4434}
(\byear{2018})
\end{barticle}
\endbibitem

\bibitem[\protect\citeauthoryear{Makantasis et~al.}{2015}]{makantasis2015deep}
\begin{bchapter}
\bauthor{\bsnm{Makantasis}, \binits{K.}},
\bauthor{\bsnm{Karantzalos}, \binits{K.}},
\bauthor{\bsnm{Doulamis}, \binits{A.}},
\bauthor{\bsnm{Doulamis}, \binits{N.}}:
\bctitle{Deep supervised learning for hyperspectral data classification through convolutional neural networks}.
In: \bbtitle{2015 IEEE International Geoscience and Remote Sensing Symposium (IGARSS)},
pp. \bfpage{4959}--\blpage{4962}
(\byear{2015}).
\bcomment{IEEE}
\end{bchapter}
\endbibitem

\end{thebibliography}

\end{document}